%% file: etacp_gammajpsi.tex
\RequirePackage{lineno}
\documentclass[prd,twocolumn,showpacs,amsmath,amssymb]{revtex4-1}
\usepackage{graphicx}
\usepackage{dcolumn}
\usepackage{bm}
\usepackage{rotating}
\usepackage{times}
\usepackage{multirow}
\usepackage{epsfig}
\usepackage{float}
\usepackage{overpic}
\usepackage{color}
\usepackage{subfigure}
\usepackage{mathrsfs}
\usepackage{booktabs}
\usepackage{relsize}

\begin{document}

\title{\boldmath Measurement of higher-order multipole amplitudes in
  $\psi(3686)\rightarrow\gamma\chi_{c1,2}$ with $\chi_{c1,2}\to\gamma J/\psi$
  and search for the  transition $\eta_{c}(2S)\to\gamma J/\psi$}

\input{author_20151124_new}

\begin{abstract}
  Using 106~million $\psi(3686)$ events collected with the BESIII
  detector, we measure multipole amplitudes for the decay
  $\psi(3686)\rightarrow\gamma\chi_{c1,2}\to\gamma\gamma J/\psi$
  beyond the dominant electric-dipole amplitudes.  The normalized
  magnetic-quadrupole amplitude for
  $\psi(3686)\rightarrow\gamma\chi_{c1,2}\rightarrow\gamma\gamma
  J/\psi$ and the normalized electric-octupole amplitudes for
  $\psi(3686)\rightarrow\gamma\chi_{c2}$,~$\chi_{c2}\rightarrow\gamma
  J/\psi$ are determined. The M2 amplitudes for
  $\psi(3686)\rightarrow\gamma\chi_{c1}$ and
  $\chi_{c1,2}\rightarrow\gamma J/\psi$ are found to differ
  significantly from zero and are consistent with theoretical
  predictions.  We also obtain the ratios of M2 contributions
  of $\psi(3686)$ and $J/\psi$ decays to $\chi_{c1,2}$,
  $b_{2}^{1}/b_{2}^{2} = 1.35\pm0.72$ and
  $a_{2}^{1}/a_{2}^{2} = 0.617\pm0.083$, which agree well with
  theoretical expectations.  By considering the multipole
  contributions of $\chi_{c1,2}$, we measure the product branching
  fractions for the cascade decays
  $\psi(3686)\rightarrow\gamma\chi_{c0,1,2}\to\gamma\gamma J/\psi$ and
  search for the process $\eta_{c}(2S)\to\gamma J/\psi$ through
  $\psi(3686)\rightarrow\gamma\eta_{c}(2S)$. The product branching
  fraction for
  $\psi(3686)\rightarrow\gamma\chi_{c0}\to\gamma\gamma J/\psi$ is
  3$\sigma$ larger than published measurements, while those of
  $\psi(3686)\rightarrow\gamma\chi_{c1,2}\to\gamma\gamma J/\psi$ are
  consistent.  No significant signal for the
  decay $\psi(3686)\to\gamma \eta_c(2S)\to\gamma \gamma J/\psi$ is observed,
  and the upper limit of the product branching fraction at the 90\%
  confidence level is determined.
\end{abstract}


\maketitle

\section{Introduction}\label{intro}

The processes $\psi(3686)\rightarrow \gamma_{1}\chi_{c1,2}$ and $\chi_{c1,2}\rightarrow\gamma_{2}J/\psi$ are dominated by electric-dipole (E1) amplitudes
but allow for higher multipole amplitudes as well, such as the magnetic-quadrupole (M2) and electric-octupole (E3) transitions.
The contributions of these higher multipole amplitudes give information on the anomalous magnetic moment $\kappa$ of the
charm quark~\cite{M2_theory2,M2_theory1} and on the admixture of $S$- and $D$-wave states~\cite{M2_theory3}.
The normalized M2 contributions for $\psi(3686)\rightarrow\gamma_{1}\chi_{c1,2}$ and $\chi_{c1,2}\rightarrow\gamma_{2}J/\psi$,
which are referred to as $b_2^{1,2}$ and $a_2^{1,2}$ with the superscript representing $\chi_{c1,2}$, are predicted to be related to the mass of the
charm quark, $m_{c}$, and $\kappa$~\cite{M2_theory2,M2_theory1,M2_theory4}. By assuming $m_{c} = 1.5\;\text{GeV}/c^{2}$ and ignoring the mixing of $S$- and $D$-wave states,
the contributions $b_{2}^{1,2}$ and $a_{2}^{1,2}$, corrected to first order in $E_{\gamma_{1,2}}/m_c$, where $E_{\gamma_{1,2}}$ is the energy of $\gamma_{1,2}$ in the rest frame of the mother charmonium state,
are predicted~\cite{M2_theory4} to be
\begin{equation}\label{theory_predict}\footnotesize
  \begin{split}
&  b_{2}^{1} = \frac{E_{\gamma_{1}}[\psi(3686)\rightarrow\gamma_{1}\chi_{c1}]}{4m_{c}}(1+\kappa) = 0.029(1+\kappa),\\
&  a_{2}^{1} = -\frac{E_{\gamma_{2}}[\chi_{c1}\rightarrow\gamma_{2}J/\psi]}{4m_{c}}(1+\kappa) = -0.065(1+\kappa),\\
&  b_{2}^{2} = \frac{3}{\sqrt{5}}\frac{E_{\gamma_{1}}[\psi(3686)\rightarrow\gamma_{1}\chi_{c2}]}{4m_{c}}(1+\kappa) =  0.029(1+\kappa),\\
&  a_{2}^{2} = -\frac{3}{\sqrt{5}}\frac{E_{\gamma_{2}}[\chi_{c2}\rightarrow\gamma_{2}J/\psi]}{4m_{c}}(1+\kappa) = -0.096(1+\kappa),
\end{split}
\end{equation}
\noindent respectively. The ratio of the M2 contributions of $\psi(3686)\rightarrow\gamma_{1}\chi_{c1}$ to $\psi(3686)\rightarrow\gamma_{1}\chi_{c2}$
($\chi_{c1}\rightarrow\gamma_{2}J/\psi$ to $\chi_{c2}\rightarrow\gamma_{2}J/\psi$)
is independent of the $m_c$ and $\kappa$ of the charm quark to first order in $E_\gamma/m_c$ and predicted to be
$b_{2}^{1}/b_{2}^{2} = 1.000\pm0.015$ and $a_{2}^{1}/a_{2}^{2} = 0.676\pm0.071$, respectively~\cite{CLEO-c}, where the dominant uncertainties come from
ignoring contributions of higher-order in $(E_\gamma/m_c)^2$.
Higher order multipole amplitudes can be obtained by investigating the angular distributions of the final-state particles~\cite{M2_theory1,angular1,angular2}.
Several experiments have searched for higher-order multipole amplitudes~\cite{crystal_ball,E-760,E-835,BESII,CLEO-c,BESIII}.
The CLEO experiment reported significant M2 contributions in
$\psi(3686)\rightarrow\gamma_{1}\chi_{c1}$ and $\chi_{c1,2}\rightarrow\gamma_{2}J/\psi$ by analyzing 24 million $\psi(3686)$ decays~\cite{CLEO-c}.
Recently, BESIII found evidence for the M2 contribution in $\psi(3686)\rightarrow\gamma\chi_{c2}$ with
$\chi_{c2}\rightarrow\pi^{+}\pi^{-}/K^{+}K^{-}$~\cite{BESIII}.

The experimentally observed charmonium states and their decay can be reproduced reasonably well by calculations based on a potential model and by
perturbative quantum chromodynamics~\cite{QM}. However, for the E1 radiative transitions of
$\psi(3686)\rightarrow\gamma_{1}\chi_{c0,1,2}$, there are significant discrepancies between different model predictions~\cite{QM_1,QM_chi1,Theory_1}
and the Particle Data Group (PDG) average~\cite{PDG}. The partial widths of $\psi(3686)\rightarrow\gamma_{1}\chi_{c0,1,2}$ are predicted to be
26, 29, and 24 keV, respectively, by using the Godfrey-Isgur model~\cite{Theory_1}, which deviate by $-(13\pm3.5)\%$, $(1.4\pm4.6)\%$, and $-(11.8\pm3.9)\%$
from the averages of experimental measurements~\cite{PDG}.


In this paper, we report on a measurement of the higher-order multipole amplitudes in the processes of
$\psi(3686)\rightarrow \gamma_{1}\chi_{c1,2},\chi_{c1,2}\rightarrow\gamma_{2}J/\psi$, where the $J/\psi$ is reconstructed in its decay modes $J/\psi\rightarrow \ell^{+}\ell^{-}~(\ell=e/\mu)$.
The measurements make use of the joint distributions of the five helicity angles in the final-state. Using the invariant mass of $\gamma_{2}J/\psi$, we obtain the product branching fractions of
$\psi(3686)\rightarrow\gamma_{1}\chi_{c0,1,2}\to\gamma_1\gamma_2 J/\psi$ and search for $\eta_{c}(2S)\to\gamma_{2}J/\psi$ produced
through $\psi(3686)\rightarrow\gamma_{1}\eta_{c}(2S)$. In the measurement of the product branching fractions of $\psi(3686)\rightarrow\gamma_{1}\chi_{c0,1,2}\to\gamma_1\gamma_2 J/\psi$,
the multipole contributions of $\chi_{c1,2}$ are considered for the first time. The results presented in this manuscript supersede the ones in Ref.~\cite{BESIII_chicj}. The analyses are based on a sample of 156~$\rm pb^{-1}$
taken at a center-of-mass energy 3.686~GeV, corresponding to 106~million $\psi(3686)$~\cite{psip_N}.
A $928$~$\rm pb^{-1}$ data sample taken at 3.773~GeV~\cite{3773_N} and a 44~pb$^{-1}$ data sample taken at 3.65~GeV are used to estimate the backgrounds from QCD processes.

\section{BESIII detector and Monte Carlo simulation}

The BESIII detector is described in detail in Ref.~\cite{BESIII_detector}. It is an approximately cylindrically symmetric detector
which covers 93\% of the solid angle around the collision point. The detector consists of four main components: (a) a 43-layer main drift chamber provides a momentum resolution of 0.5\% for charged tracks at 1~GeV/$c$ in a 1~T magnetic field; (b) a time-of-flight system (TOF) is
constructed of plastic scintillators with a time resolution of 80~ps (110~ps) in the barrel (end caps); (c) a 6240 cell CsI(Tl) crystal electromagnetic
calorimeter (EMC) provides an energy resolution for photons of 3.0\% (5.0\%) around 0.3~GeV in the barrel (end caps)~\cite{photon_resolution};  (d) a muon counter consisting of
nine (eight) layers of resistive plate chambers in the barrel (end caps) within the return yoke of the magnet with a position resolution of 2~cm provides
muon/pion separation. A {\sc geant4}~\cite{GEANT4} based detector simulation package has been developed to model the detector response used in
Monte Carlo (MC) generated events.

A MC simulated sample of 106~million generic $\psi(3686)$ decays ("inclusive MC") is used for general background studies. The
$\psi(3686)$ resonances are produced by the event generator {\sc Kkmc}~\cite{KKMC}. The known decays are generated by {\sc BesEvtGen}~\cite{EvtGen} with
branching fractions taken from the PDG~\cite{PDG}, while the remaining decays are generated according to the {\sc Lundcharm} model~\cite{LundCharm}.
Exclusive MC samples for signal decays are generated to optimize the selection criteria and to determine the detection efficiencies.
The $\psi(3686)\rightarrow\gamma\chi_{c0,1,2}\rightarrow\gamma\gamma J/\psi$ decays are generated with angular distributions determined
from data, and the $\eta_{c}(2S)\rightarrow\gamma J/\psi$ decay is generated according to the {\sc Helamp} model in {\sc EvtGen}~\cite{EvtGen}.
To estimate the background contributions from $\psi(3686)$ decays, the exclusive MC samples $\psi(3686)\to\eta J/\psi, \pi^0 J/\psi,\pi^0\pi^0 J/\psi, \gamma\gamma J/\psi$ are generated according to the {\sc Helamp}, {\sc Jpipi}~\cite{EvtGen}, and {\sc Phsp} models, respectively. To investigate QED processes backgrounds, radiative Bhabha and dimuon events ($e^+e^-\to e^+e^-/\mu^+\mu^-$) simulated with {\sc Babayaga V3.5}~\cite{babayaga}, as well as $\psi(3770)\to\gamma\chi_{cJ}$ and $\gamma_{\rm ISR} \psi(3686)\to\gamma\chi_{cJ},\pi^0 J/\psi$ produced by {\sc Kkmc}~\cite{KKMC}, are used together with the experimental data at 3.773 GeV.

\section{Event selection}

The signal decay $\psi(3686)\rightarrow\gamma_{1}\chi_{c0,1,2}(\eta_{c}(2S))\rightarrow\gamma_1\gamma_{2}J/\psi,~J/\psi\rightarrow \ell^{+}\ell^{-} ~(\ell=e,\mu)$
consists of two charged tracks and two photons. Events with exactly two oppositely charged tracks and from two up to four photon candidates are selected.
Charged tracks are required to originate from the run-dependent interaction point within $1\rm~cm$ in the direction perpendicular to
and within $\pm10\rm~cm$ along the beam axis and should lie within the polar angular region of $|\cos\theta|<0.93$.
The momentum $p$ of each track must be larger than 1~GeV/$c$. The energy deposit $E$ in the EMC and $E/p$ of each track are used to identify muon or electron candidates.
Tracks with $E < 0.4$ GeV are taken as muons, and those with $E/p > 0.8~c$ are identified as electrons. Events with both tracks identified as muons or electrons are accepted for further analysis. Photons are reconstructed from isolated showers in the EMC, where the angle between the positions in the EMC of the
photon and the closest charged track is required to be larger than $10$ deg. The energy deposited in the EMC is corrected by the energy loss in nearby TOF counters to improve the reconstruction efficiency and the
energy resolution.  The energy of each photon shower is required to be larger than 25~MeV.
The shower timing information is required to be in coincidence with the event start time with a requirement of $0 \leq t \leq 700$~ns to suppress electronic noise and showers unrelated to the event.

A four-constraint (4C) kinematic fit is performed for the two lepton candidates and all possible two photon combinations with the
initial $\psi(3686)$ 4-momentum as a constraint. If more than one combination is found in one event, the one with the smallest $\chi^{2}_{4\rm C}$ value is kept.
The $\chi^{2}_{4\rm C}$ is required to be $\chi^{2}_{4\rm C} < 60$, where the requirement is determined by optimizing the statistical significance $S/\sqrt{S+B}$
for the $\eta_{c}(2S)$ channel.  Here, $S$ is the number of events in the $\eta_{c}(2S)$ signal region
$3.60 < M^{\rm 4C}(\gamma_{2}\ell^{+}\ell^{-}) < 3.66$ GeV/$c^{2}$ ($\gamma_{2}$ denotes the photon with larger energy, and $M^{\rm 4C}$ is the invariant mass with the energies and momenta updated with the 4C kinematic fit) obtained from the exclusive MC sample, and $B$ is the number of corresponding background events determined
from the 106~million inclusive MC sample and a continuum data sample collected at a center-of-mass energy of 3.65~GeV. The latter is normalized to
the luminosity of the $\psi(3686)$ data sample. The branching fraction of the decay $\eta_{c}(2S)\rightarrow\gamma_{2}J/\psi$ is assumed to be 1\%.

To select events including the $J/\psi$ intermediate state, the invariant mass of the lepton pair is required to be in the region of
$3.08 < M^{4\rm C}(\ell^{+}\ell^{-}) < 3.12$ GeV/$c^{2}$. In addition, to remove $\psi(3686)\to\pi^{0}J/\psi$ and $\psi(3686)\to\eta J/\psi$ backgrounds,
events with an invariant mass of the photon pair in the regions $0.11 < M^{4\rm C}(\gamma\gamma) < 0.15$~GeV/$c^{2}$ or
$M^{4\rm C}(\gamma\gamma) > 0.51$~GeV/$c^{2}$ are rejected. A MC study shows that this removes 97.9\% of the $\pi^{0}J/\psi$ events
and almost 100\% of the $\eta J/\psi$ events, while the efficiencies of the signal channels for $\chi_{c0},\chi_{c1},\chi_{c2}$, and $\eta_{c}(2S)$
are 74.7\%, 90.0\%, 93.9\%, and 88.0\%, respectively.

\section{Measurement of higher-order multipole amplitudes}\label{higher_multipole}

\begin{figure}
  \centering
    \includegraphics[width=0.4\textwidth]{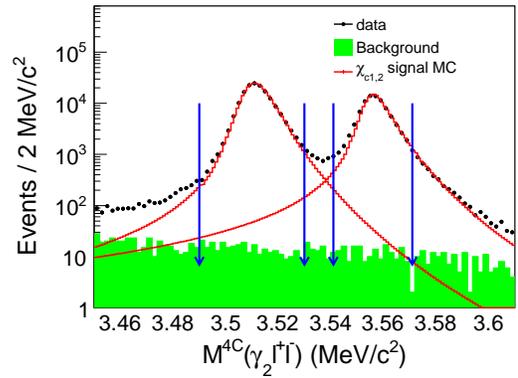}
   \caption{Mass distributions of $M^{\rm 4C}(\gamma_2 \ell^+\ell^-)$ for events in the $\chi_{c1,2}$ region. Black dots correspond to data, and
red histograms are obtained from the signal MC samples scaled by the maximum bin. The green dashed histogram is the background contribution obtained from the inclusive MC samples.
The arrows denote the signal regions.}
  \label{chicj_dis} 
\end{figure}

Figure~\ref{chicj_dis} shows the $M^{4\rm C}(\gamma_{2}\ell^{+}\ell^{-})$ invariant-mass distribution for the selected $\chi_{c1,2}$ candidates.
The signal regions for $\chi_{c1}$ and $\chi_{c2}$ are defined as 3.496~$ < M^{4\rm C}(\gamma_{2}\ell^+\ell^-) < $3.533~GeV/$c^2$ and
3.543~$ < M^{4\rm C}(\gamma_{2}\ell^+\ell^-) <$ 3.575~GeV/$c^2$, respectively. We find 163922 $\chi_{c1}$ candidates and 89409 $\chi_{c2}$ candidates. The background is estimated from the inclusive MC sample.  The total number of background events is found to be 1016 (0.7\%)
within the $\chi_{c1}$ signal region and 883 (1.0\%) in the $\chi_{c2}$ region. For the $\chi_{c1}$ ($\chi_{c2}$) channel, the dominant background
is the contamination from $\chi_{c2}$ ($\chi_{c1}$). Some backgrounds stem from $\psi(3686)\rightarrow\gamma\gamma J/\psi$ and $\pi^{0}\pi^{0}J/\psi$ decays. The QED process $e^+e^-\to \ell^+\ell^-\gamma_{\rm ISR/FSR}$ contributes about 109 events for $\chi_{c1}$ and 135 events for $\chi_{c2}$. Non-$J/\psi$ background is negligibly small according to the sideband analysis.

Events in the signal regions are used to determine the higher-order multipole amplitudes in the $\psi(3686)\rightarrow\gamma_{1}\chi_{c1,2}\rightarrow\gamma_1\gamma_{2}J/\psi$ radiative transitions.
The normalized M2 contributions for the channels $\psi(3686)\rightarrow\gamma_{1}\chi_{c1,2}$ and $\chi_{c1,2}\rightarrow\gamma_{2}J/\psi$ are denoted
as $b_{2}^{1,2}$ and $a_{2}^{1,2}$, respectively. In the $\chi_{c2}$ decays, the E3 transition is also allowed. The corresponding normalized E3 amplitudes are indicated
as $b^{2}_{3}$ and $a^{2}_{3}$ for $\psi(3686)\rightarrow\gamma_{1}\chi_{c2}$ and $\chi_{c2}\rightarrow\gamma_{2}J/\psi$, respectively.

\subsection{Fit method}

We perform an unbinned maximum likelihood fit to obtain the higher-order multipole amplitudes following the procedure as described in Ref.~\cite{BESIII}.
The log-likelihood function is built as $\ln\mathcal{L}_{s} = \ln\mathcal{L}-\ln\mathcal{L}_{b}$, where $\mathcal{L} \equiv \prod_{i=1}^{N}{\rm F}_{\chi_{c1,2}}(i)$
denotes the product of probability densities for all candidates in the signal region, $N$ is the number of the candidates, and $F$ is the probability density functions (PDFs).  The contribution to the likelihood from background events, $\mathcal{L}_{b}$, is estimated using the inclusive MC sample and continuum data.

The PDFs $F$ for the joint angular distributions of the $\chi_{c1,2}$ decay sequences are defined as
$\frac{ W_{\chi_{cJ}}(\theta_{1},\theta_{2},\phi_{2},\theta_{3},\phi_{3},a_{2,3}^{J},b_{2,3}^{J})}{\overline{W_{\chi_{cJ}}(a_{2,3}^{J},b_{2,3}^{J})}}$.
The term in the numerator, $W_{\chi_{cJ}}(\theta_{1},\theta_{2},\phi_{2},\theta_{3},\phi_{3},a_{2,3}^{J},b_{2,3}^{J})$, is derived from the helicity amplitudes
and the Clebsch-Gordan relation~\cite{M2_theory1}, while $\overline{W_{\chi_{cJ}}(a_{2,3}^{J},b_{2,3}^{J})}$ is used for the normalization.
$\theta_{1}$ is the polar angle of $\gamma_1$ in the $\psi(3686)$ rest frame with the $z$ axis in the electron-beam direction.
$\theta_{2}$ and $\phi_{2}$ are the polar and azimuthal angles of $\gamma_{2}$ in the $\chi_{cJ}$ rest frame with the $z$ axis in the $\gamma_{1}$ direction
and $\phi_{2}=0$ in the electron-beam direction. $\theta_{3}$ and $\phi_{3}$ are the polar and azimuthal angles of $\ell^{+}$ from
$J/\psi\rightarrow \ell^{+}\ell^{-}$ in the $J/\psi$ rest frame with the $z$ axis aligned to the $\gamma_{2}$ direction and $\phi_{3}=0$ in the $\gamma_{1}$ direction.

The formula $W_{\chi_{cJ}}(\theta_{1},\theta_{2},\phi_{2},\theta_{3},\phi_{3},a_{2,3}^{J},b_{2,3}^{J})$ for the helicity amplitudes has been discussed
in Refs.~\cite{BESII,CLEO-c,BESIII,amp_012}. Using the same method as reported in Refs.~\cite{BESII,CLEO-c,BESIII}, the joint angular distributions
$W_{\chi_{cJ}}$ can be expressed in terms of $a_{2,3}^{J}$ and $b_{2,3}^{J}$ as

\begin{equation}\label{chic1_func}
\begin{split}
 &W_{\chi_{cJ}}(\theta_{1},\theta_{2},\phi_{2},\theta_{3},\phi_{3},a_{2,3}^{J},b_{2,3}^{J}) \\
 =& \sum_{n}a_{n}A_{|\nu|}^{J}A_{|\widetilde{\nu}|}^{J}B_{|\nu'|}^{J}B_{|\widetilde{\nu}'|}^{J},
\end{split}
\end{equation}

\noindent with

\begin{equation}
\begin{split}
&\begin{pmatrix}
A_0^{1}\\
A_1^{1}
\end{pmatrix}
=\begin{pmatrix}
\sqrt{0.5} & \sqrt{0.5} \\
\sqrt{0.5} & -\sqrt{0.5}
\end{pmatrix}
  \begin{pmatrix}
  a_1^{1} \\
  a_2^{1}
\end{pmatrix}, \\
&\begin{pmatrix}
B_0^{1}\\
B_1^{1}
\end{pmatrix}
=\begin{pmatrix}
\sqrt{0.5} & \sqrt{0.5} \\
\sqrt{0.5} & -\sqrt{0.5}
\end{pmatrix}
  \begin{pmatrix}
  b_1^{1} \\
  b_2^{1}
\end{pmatrix}, \\
& \begin{pmatrix}
  A_0^{2} \\
  A_1^{2} \\
  A_2^{2}
\end{pmatrix}
=\frac{1}{\sqrt{30}}\begin{pmatrix}
\sqrt{3} & \sqrt{15} & 2\sqrt{3} \\
3 & \sqrt{5} & -4 \\
3\sqrt{2} & -\sqrt{10} & \sqrt{2}
\end{pmatrix}
  \begin{pmatrix}
  a_1^{2} \\
  a_2^{2} \\
  a_3^{2}
\end{pmatrix},\\
&\begin{pmatrix}
  B_0^{2} \\
  B_1^{2} \\
  B_2^{2}
\end{pmatrix}
=\frac{1}{\sqrt{30}}\begin{pmatrix}
\sqrt{3} & \sqrt{15} & 2\sqrt{3} \\
3 & \sqrt{5} & -4 \\
3\sqrt{2} & -\sqrt{10} & \sqrt{2}
\end{pmatrix}
  \begin{pmatrix}
  b_1^{2} \\
  b_2^{2} \\
  b_3^{2}
\end{pmatrix},\\
\end{split}
\end{equation}

\noindent where $B_{|\nu|}^{J}$ and $B_{|\widetilde{\nu}|}^{J}$~\cite{amp_012} are the helicity amplitudes for $\psi(3686)\rightarrow\gamma_{1}\chi_{cJ}$, $A_{|\nu|}^{J}$ and $A_{|\widetilde{\nu}|}^{J}$~\cite{amp_012} are those for $\chi_{cJ}\rightarrow\gamma_{2} J/\psi$.
$\sqrt{(a_1^{1})^2+(a_2^{1})^2}=1$, $\sqrt{(a_1^{2})^2+(a_2^{2})^2+(a_3^{2})^2}=1$, and similarly for $b^J_{|\nu|}$s.
The coefficients $a_{n}(^{n=1,\ldots,9 ~{\rm for}~ \chi_{c1}}_{n=1,\ldots,36 ~{\rm for} ~\chi_{c2}})$ are functions of $\theta_{1},\theta_{2},\phi_{2},\theta_{3},\phi_{3}$.
For the normalization, high-statistics phase-space ({\sc PHSP}) MC samples are generated.

The normalization factor is expressed as
\begin{equation}\label{normal_func}
  \begin{split}
 &\overline{W_{\chi_{cJ}}(a_{2,3}^{J},b_{2,3}^{J})} \\
= &\frac{\sum_{i=1}^{N_P}W_{\chi_{cJ}}(\theta_{1}(i),\theta_{2}(i),\phi_{2}(i),\theta_{3}(i),\phi_{3}(i),a_{2,3}^{J},b_{2,3}^{J})}{N_P} \\
=& \sum_{n}\overline{a_{n}}A_{|\nu|}^{J}A_{|\widetilde{\nu}|}^{J}B_{|\nu'|}^{J}B_{|\widetilde{\nu}'|}^{J},
\end{split}
\end{equation}
\noindent where $N_{P}$ is the number of selected events. In such a way, the detector efficiency is considered in the normalization.

\subsection{Fit results}

\begin{figure*}
  \centering
  \includegraphics[width=0.95\textwidth]{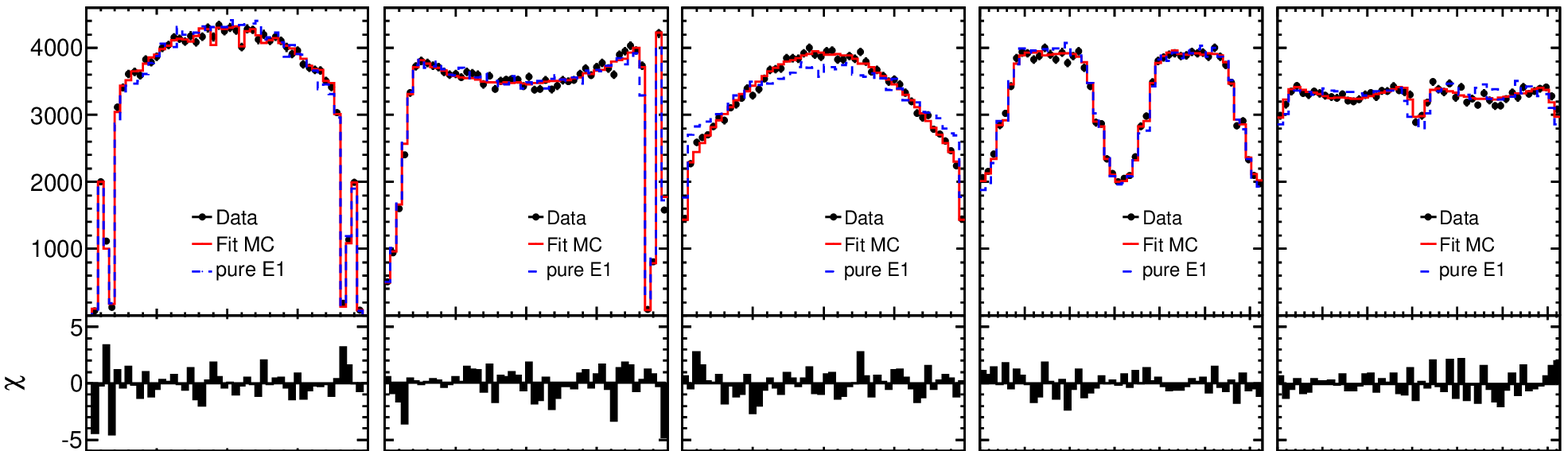}\\
  \includegraphics[width=0.95\textwidth]{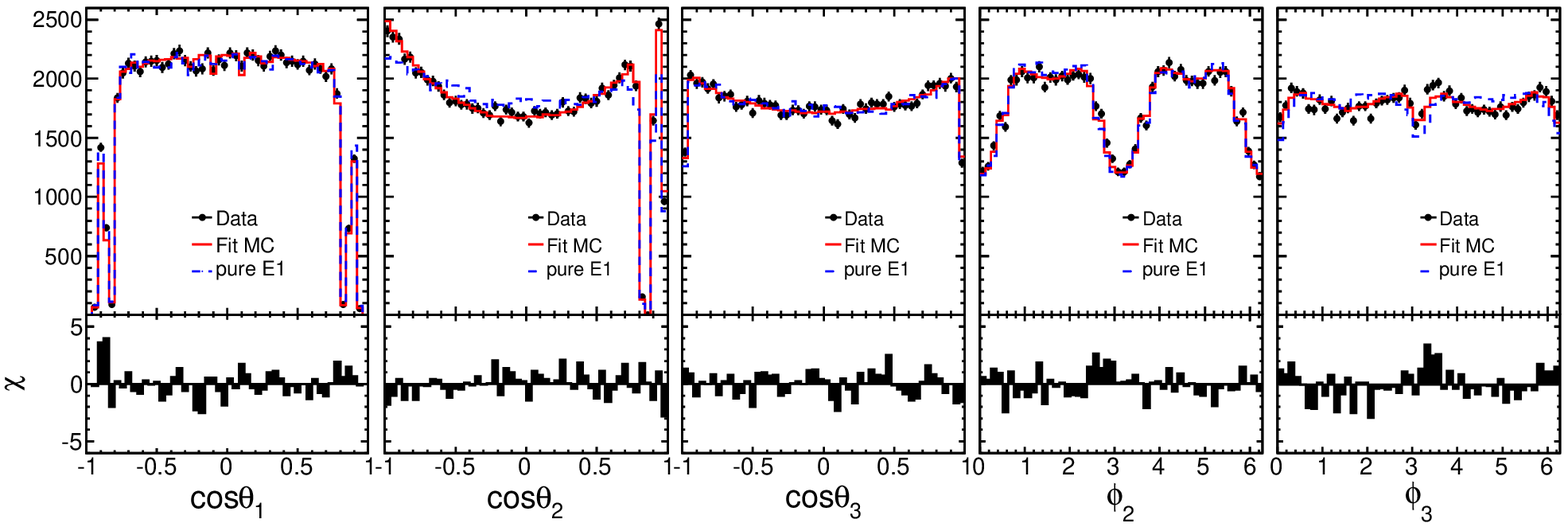}\\
  \caption{Results of the multidimensional fit on the joint angular distribution and the projections on
$\cos\theta_{1}$, $\cos\theta_{2}$,  $\cos\theta_{3}$, $\phi_{2}$, $\phi_{3}$ of the final-state particles.
The upper ten plots show the angular distributions for the $\chi_{c1}$ channel, and the lower ones are for the $\chi_{c2}$ channel.
The black dots with error bars represent data subtracted by background, the red histograms are the fit results, and the blue dashed lines are pure E1 distributions. The lower plots depict the relative residual $\chi = (N_{\rm data}-N_{\rm fit})/\sqrt{N_{\rm data}}$ of the fit.}\label{chic1_data}
\end{figure*}

By minimizing $-\ln\mathcal{L}_{s}$, the best estimates of the high-order multipole amplitudes can be obtained.
To validate the fit procedure, checks are performed with MC samples for $\chi_{c1,2}$ separately, where the MC samples are generated based on a pure E1 transition model ($a_{2,3}^{1,2}=0,~b_{2,3}^{1,2}=0$) or an arbitrary higher-order multipole amplitude ($a_{2,3}^{1,2}\neq0,~b_{2,3}^{1,2}\neq0$).
The fit values are consistent with the input values within 1$\sigma$ of statistical uncertainty.
An unbinned maximum likelihood fit to the joint angular distribution for data is performed, and the corresponding angular distributions are depicted
in Fig.~\ref{chic1_data} together with the relative residual spectra. The fit results are listed in Table~\ref{chic1_result_sys}, where the first uncertainties are statistical and the second ones are systematical as described in Sec.~\ref{sys_error_section}.

\begin{table*}[!htbp]
  \centering
  \caption{\label{chic1_result_sys}Fit results for $a_{2,3}^J$ and $b_{2,3}^J$ for the process of $\psi(3686)\rightarrow\gamma_{1}\chi_{c1,2}\rightarrow\gamma_{1}\gamma_{2}J/\psi$; the first uncertainty is statistical, and the second is systematic. The $\rho_{a_{2,3}b_{2,3}}^{J}$ are the correlation coefficients between $a_{2,3}^J$ and $b_{2,3}^J$.}
\begin{tabular}{c|cc}
\hline
  \multirow{2}{*}{$\chi_{c1}$}     &   $a_2^{1}=-0.0740\pm0.0033\pm0.0034$, $b_2^{1}=0.0229\pm0.0039\pm0.0027$ &\\[0.05in]
                                       &   $\rho_{a_2b_2}^{1} = 0.133$  &\\[0.05in]
\hline
  \multirow{4}{0.2in}{$\chi_{c2}$}     &   $a_2^{2}=-0.120\pm0.013\pm0.004$, $b_2^{2}=0.017\pm0.008\pm0.002$ &\\[0.05in]
       &   $a_3^{2}=-0.013\pm0.009\pm0.004$, $b_3^{2}=-0.014\pm0.007\pm0.004$ &\\[0.05in]
       &   $\rho_{a_2b_2}^{2} = -0.605$, $\rho_{a_2a_3}^{2} = 0.733$, $\rho_{a_2b_3}^{2} = -0.095$ &\\[0.05in]
       &   $\rho_{a_3b_2}^{2} = -0.422$, $\rho_{b_2b_3}^{2} = 0.384$, $\rho_{a_3b_3}^{2} = -0.024$ &\\[0.05in]
\hline
\end{tabular}
\end{table*}

The statistical significance of a nonpure E1 transition is calculated to be
24.5$\sigma$ (13.5$\sigma$) for $\chi_{c1}$ ($\chi_{c2}$) by taking the difference of the log-likelihood values for the fits with
higher-order multipole amplitudes included and fits based on a pure E1 transition, taking the change in the number of degrees of freedom, $\Delta ndf = 2~(4)$,
into consideration. Similarly, the statistical significance of the E3 contribution for $\chi_{c2}$ is 2.3$\sigma$, as obtained by comparing the log-likelihood values between the nominal fit and a fit based on the assumption that E3 contribution is zero.
A Pearson-$\chi^{2}$ test~\cite{Pearson_test} is performed to validate the fit result. Each angular dimension
(i.e., $\cos\theta_{1}, \cos\theta_{2}, \phi_{2}, \cos\theta_{3}, \phi_{3}$) is divided equally into eight bins.
This leads to a total of $8^5 = 32768$ cells. The $\chi^{2}$ is defined as
\begin{equation}\label{chi2_define}
  \chi^{2} = \mathop{\sum_{i}}\frac{(n_{i}^{\rm DT}-n_i^{\rm BKG}-n_{i}^{\rm MC})^{2}}{n_{i}^{\rm DT}+n_i^{\rm BKG}},
\end{equation}
\noindent where $n_{i}^{\rm DT}$ is the number of events in the $i$th cell for data, $n_i^{\rm BKG}$ is the number of the background contribution determined by the inclusive MC sample, and
$n_{i}^{\rm MC}$ is the number of events for the luminosity-normalized MC sample produced according to the best fit values for $a_{2,3}^J$ and $b_{2,3}^J$.
The number of events of the MC sample is 40 times larger than of the data. For cells with fewer than ten events,
events in adjacent bins are combined.
The test results in $\chi^{2}/ndf = 9714.7/9563 = 1.02$ for $\chi_{c1}$ and $\chi^{2}/ndf = 5985.2/5840 = 1.02$ for $\chi_{c2}$, demonstrating that the fit gives an excellent representation of the data.

\section{\boldmath{Measurement of $\mathcal{B}(\psi(3686)\to\gamma\chi_{cJ}\to\gamma\gamma J/\psi)$ and search for the process} $\eta_{c}(2S)\to\gamma J/\psi$ }

With the selected $e^{+}e^{-}\rightarrow\gamma_{1}\gamma_{2}J/\psi$ candidates, we measure the product branching fractions of the decay
$\psi(3686)\rightarrow\gamma_{1}\chi_{c0,1,2}\to\gamma_1\gamma_2 J/\psi$ and search for the process $\eta_{c}(2S)\rightarrow\gamma_{2}J/\psi$.
For the $J/\psi\rightarrow e^{+}e^{-}$ channel, additional requirements are applied to suppress the background from
radiative Bhabha events [$e^{+}e^{-}\rightarrow\gamma_{\rm ISR/FSR} e^{+}e^{-}$, where $\gamma_{\rm ISR/FSR}$ denotes the initial-/final-state radiative (ISR/FSR) photon(s)].
Since the electron (positron) from radiative Bhabha tends to have a polar angle $\cos\theta_{e^{+}(e^{-})}$ close to +1 (-1), we
apply a requirement of $\cos\theta_{e^{+}} < 0.3$ and $\cos\theta_{e^{-}} > -0.3$.
These requirements suppress 77\% of the Bhabha events with a reduction of the signal efficiency by one-third.
The corresponding MC-determined efficiencies are listed in Table~\ref{eff}.

\begin{table*}[!htbp]
\centering
  \caption{\label{eff}Detection efficiencies ($\epsilon$) for channels of $\psi(3686)\to\gamma\chi_{c0,1,2},\gamma\eta_c(2S),\gamma\gamma J/\psi$ and
  the number ($N$) of estimated background for channels $\psi(3686)\to\pi^0 J/\psi, \pi^0\pi^0 J/\psi$ scaled by the decay branching fraction and the total $\psi(3686)$ number.
  }
\begin{tabular}{lcccccccc}
\hline
Channel          & $\epsilon_{\chi_{c0}}$ (\%) & $\epsilon_{\chi_{c1}}$ (\%) & $\epsilon_{\chi_{c2}}$ (\%) & $\epsilon_{\eta_{c}(2S)} (\%)$ & $\epsilon_{\gamma\gamma J/\psi}$ (\%)& $N_{\pi^{0}J/\psi}$ && $N_{\pi^{0}\pi^{0}J/\psi}$\\
\hline
$e^{+}e^{-}$     &  15.1    &  20.1    &  20.3     &  16.9    & 17.1 & 26.8$\pm$0.7  && 246.5$\pm$4.5 \\
$\mu^{+}\mu^{-}$ &  32.7    &  44.1    &  44.0     &  37.0    & 38.0 & 65.2$\pm$1.7  && 500.9$\pm$9.1 \\
\hline
\end{tabular}
\end{table*}

A 4C kinematic fit has the defect that the energy of a fake and soft photon will be modified according to the topology of a signal event due to relatively large uncertainty, which results in a
peaking background signature in the $M^{\rm 4C}(\gamma_{2}J/\psi)$ invariant-mass spectrum.
To remove the peaking background, such as radiative Bhabha and radiative dimuon ($e^+e^-\to\gamma_{\rm ISR/FSR}\mu^+\mu^-$), a three-constraint (3C) kinematic fit is applied, in which
the energy of the soft photon ($\gamma_1$) is left free in the fit.
The detailed MC studies indicate that the 3C kinematic fit does not change the peak position of the invariant mass for signals and the corresponding resolutions are similar to those with the 4C kinematic fit.

\subsection{Background study}\label{background}

The backgrounds mainly come from $\psi(3686)$ transitions to $J/\psi$ and from $e^{+}e^{-}\rightarrow \ell^{+}\ell^{-} n\gamma_{\rm ISR/FSR} (\ell=e/\mu)$. The other background, including $\psi(3686)\to\eta J/\psi$, $\gamma_{\rm ISR}J/\psi$ and non-$J/\psi$ backgrounds, is only 0.3\% of that from $\psi(3686)$, which is neglected.

The backgrounds from $\psi(3686)$ transitions to $J/\psi$ include $\psi(3686)\to\gamma\gamma J/\psi, \pi^{0}\pi^{0}J/\psi,\pi^{0}J/\psi$.
High-statistics MC samples of these decays are generated to determine their distributions and contributions.
With the published branching fractions~\cite{PDG}, which have been measured precisely by different experiments,
the estimated number of events for $\psi(3686)\rightarrow\pi^{0}\pi^{0}J/\psi, \pi^{0}J/\psi$ and the efficiency
for $\psi(3686)\rightarrow\gamma\gamma J/\psi$ are obtained as summarized in Table~\ref{eff}.

The second major source of background includes radiative Bhabha and dimuon processes, $e^{+}e^{-}\rightarrow \ell^{+}\ell^{-}\gamma_{\rm ISR/FSR}(\gamma_{\rm ISR/FSR})$
and $\psi(3686)\rightarrow \ell^{+}\ell^{-}\gamma_{\rm FSR}(\gamma_{\rm FSR})~ (l=e/\mu)$. To precisely describe the shape, the background is divided up into
two parts: $\ell^{+}\ell^{-}$ with one radiative photon and $\ell^{+}\ell^{-}$ with two radiative photons. For the background from $\psi(3686)\to \ell^{+}\ell^{-}\gamma_{\rm FSR}(\gamma_{\rm FSR})$, the ratio of event yields between the two parts ($N_{\ell^+\ell^-\gamma\gamma}/N_{\ell^+\ell^-\gamma}$) is obtained by a MC simulation. For the background from radiative Bhabha/dimuon processes,
the ratio $N_{\ell^+\ell^-\gamma\gamma}/N_{\ell^+\ell^-\gamma}$ is obtained by a fit to a 928~pb$^{-1}$ data sample taken at a center-of-mass energy of 3.773 GeV.
After the event selection imposed on the data, the remaining events are mainly radiative Bhabha/dimuon events, and a small contribution originates
from $\psi(3770)\to\gamma\chi_{cJ}$ and decays of $\psi(3686)$ produced in the ISR process.
In the fit, the shapes of the $M^{3\rm C}(\gamma_{2}\ell^+\ell^-)$ distributions for the Bhabha/dimuon processes are determined
from a $\psi(3686)\rightarrow \ell^{+}\ell^{-}\gamma_{\rm FSR}(\gamma_{\rm FSR})$ MC sample by shifting the $M^{\rm 3C}(\gamma_2 \ell^{+}\ell^{-})$ from $\psi(3686)$ to $\psi(3770)$ according to the formula $m'=a*(m-m_{0})+m_{0}$,
where $m_{0} = 3.097$ GeV/$c^2$ is the mass threshold of $\gamma J/\psi$, and the coefficient $a = (3.773-m_{0})/(3.686-m_{0}) = 1.15$ shifts the events from 3.686 to 3.773 GeV.
 The shapes of the backgrounds are based on
MC simulation, while the amplitude of each component is set as a free parameter.
Thus, the cross section weighted ratio
of the backgrounds $e^+e^-\to \ell^+\ell^-\gamma_{\rm ISR/FSR}(\gamma_{\rm ISR/FSR})$ and $\psi(3686)\to \ell^+\ell^-\gamma_{\rm FSR}(\gamma_{\rm FSR})$ for the two parts is $N_{e^{+}e^{-}\gamma\gamma}/N_{e^{+}e^{-}\gamma} = 1.203\pm0.081$ ($N_{\mu^{+}\mu^{-}\gamma\gamma}/N_{\mu^{+}\mu^{-}\gamma} = 0.689\pm0.044$) for the $e^{+}e^{-}$ ($\mu^{+}\mu^{-}$) channel. The quantitative results and shapes will be used in the simultaneous fit.

\subsection{\boldmath Simultaneous fit to $M^{\rm 3C}(\gamma_{2}\ell^{+}\ell^{-})$}

\begin{figure*}[!htbp]
  \centering
   \includegraphics[width=0.4\textwidth]{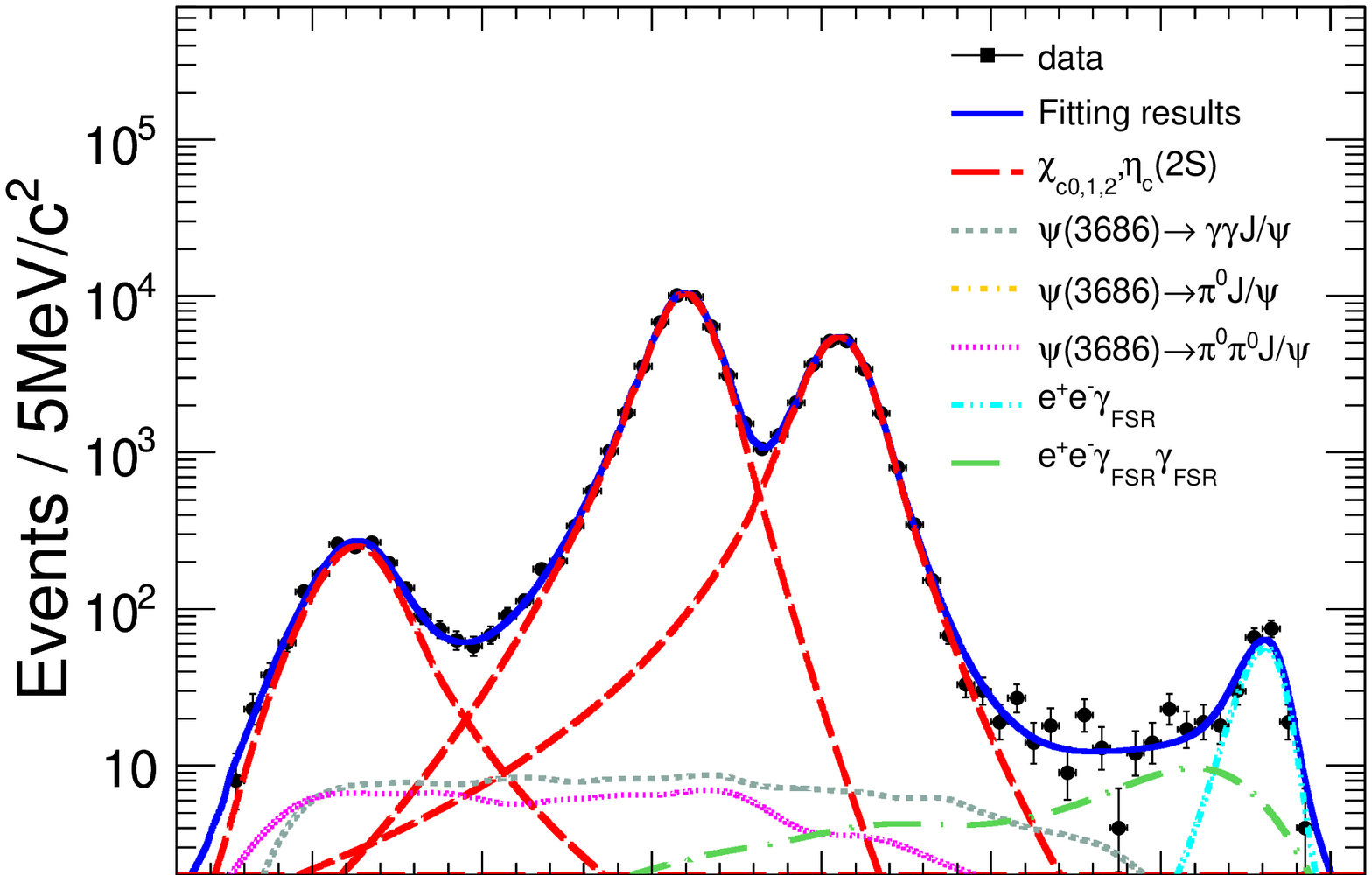}
   \includegraphics[width=0.4\textwidth]{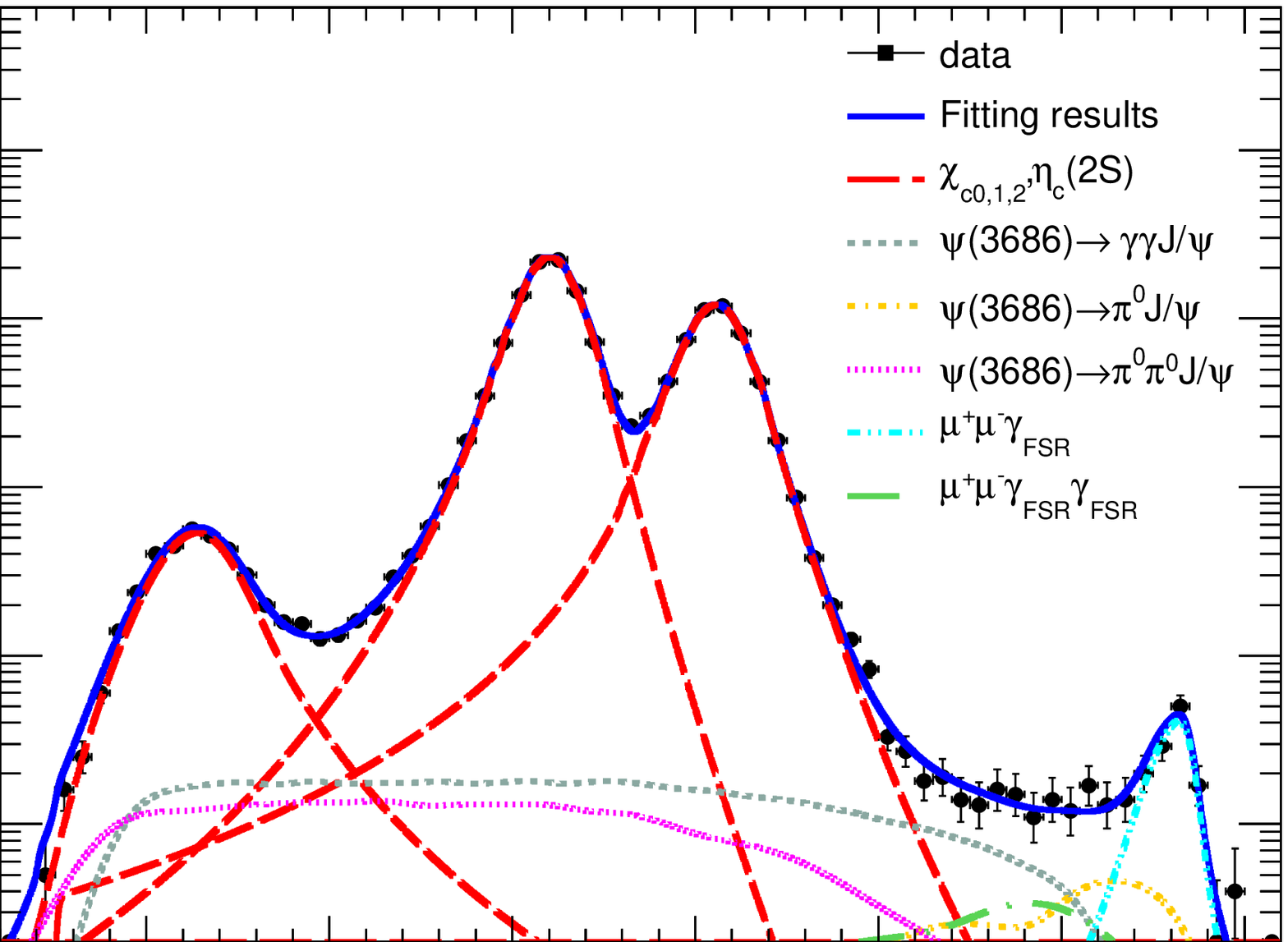}
   \includegraphics[width=0.4\textwidth]{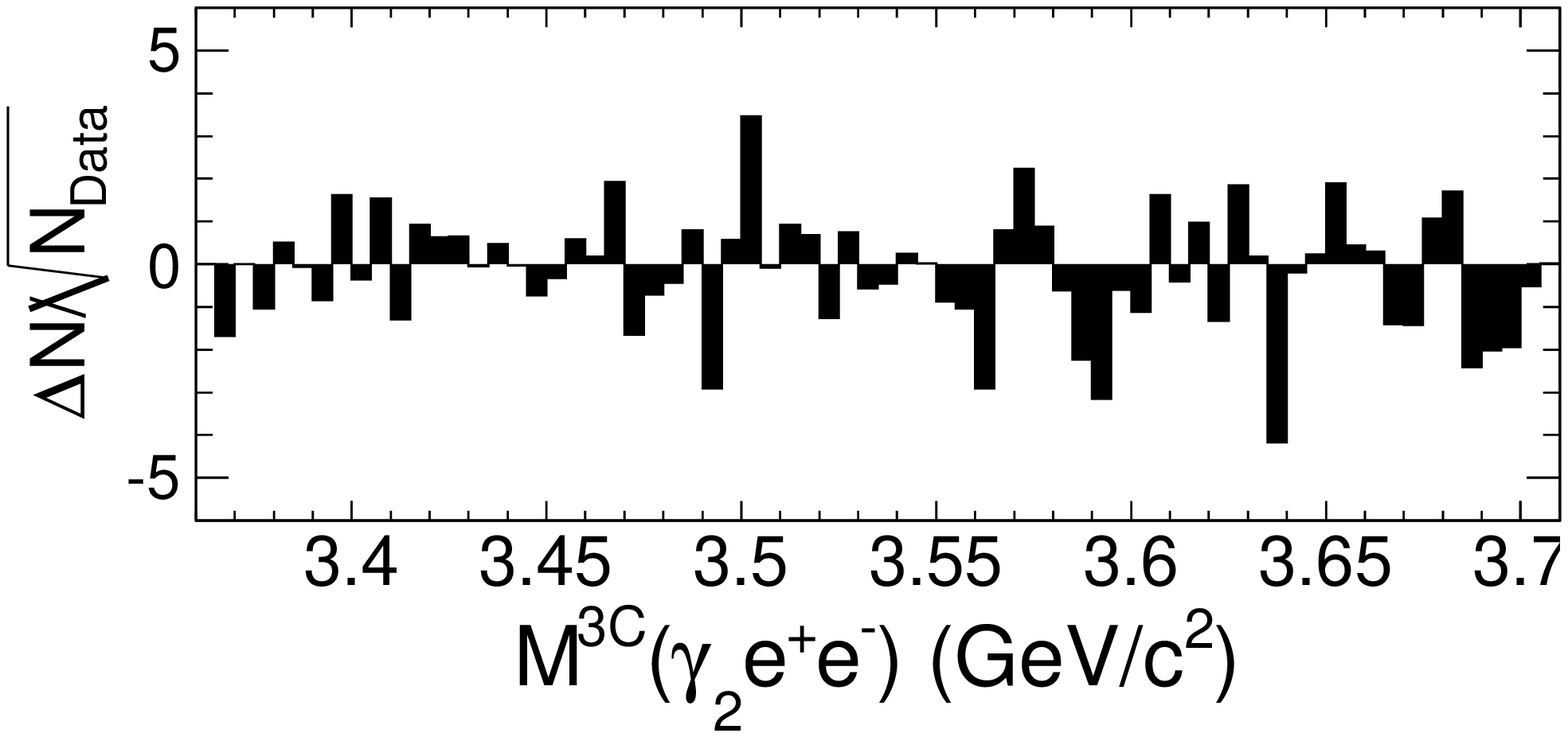}
   \includegraphics[width=0.4\textwidth]{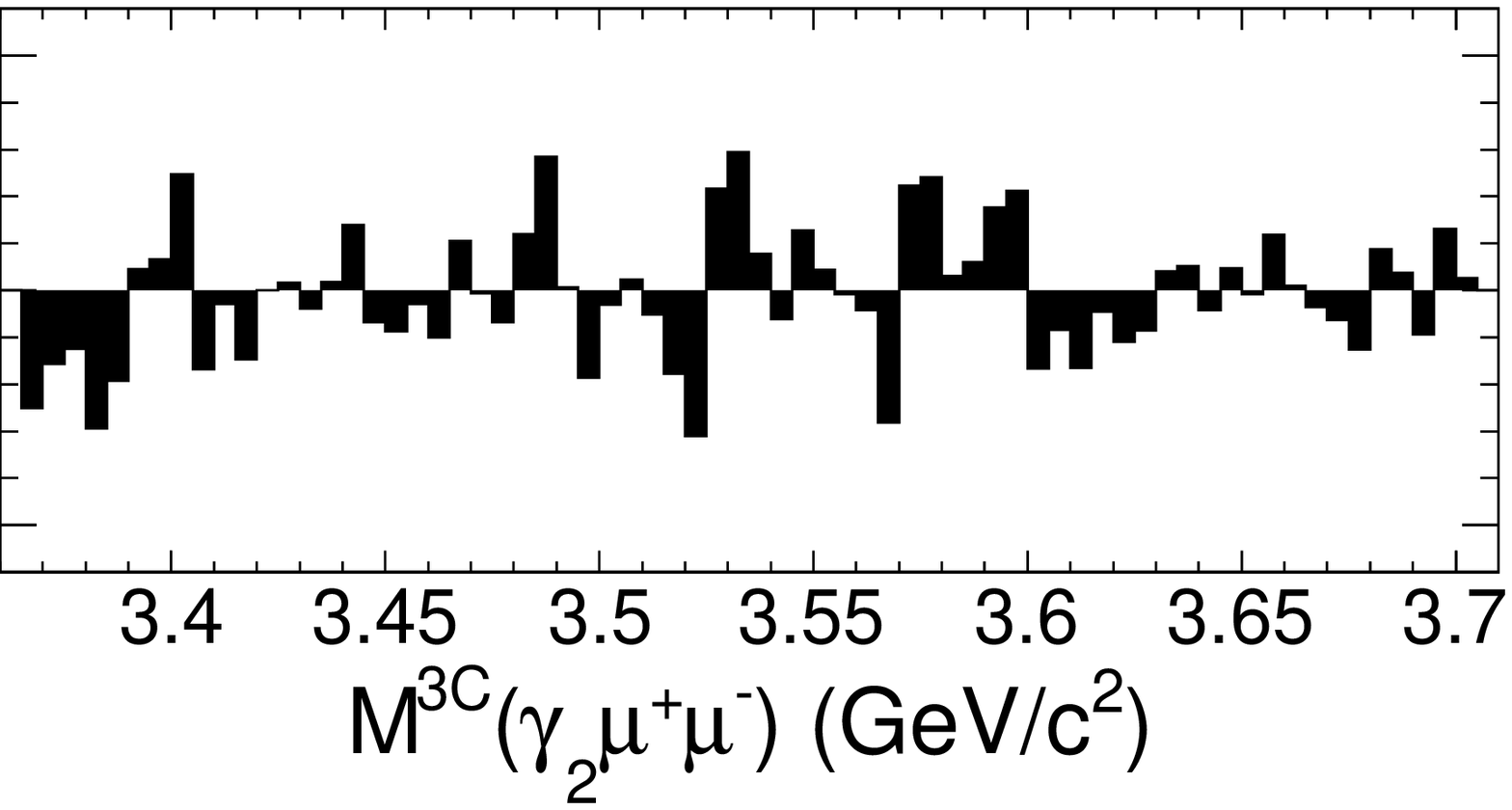}
  \caption{The results of a simultaneous maximum likelihood fit (top) and corresponding relative residual $(N_{\rm data}-N_{\rm fit})/\sqrt{N_{\rm data}}$ (bottom). The left panel is for the $e^+e^-$ channel, while right one is for the $\mu^+\mu^-$ channel. The black dots are the data, the blue curves are the fit results, and the red long-dashed lines are for $\chi_c0,1,2$ signals. The gray dashed, orange dot-dashed, and pink dotted lines are for backgrounds of $\psi(3686)\to\gamma\gamma J/\psi,~ \pi^0 J/\psi, {\rm and} ~ \pi^0\pi^0 J/\psi$, respectively. The light-blue dot-dot-dashed and green dot-long-dashed lines are for backgrounds with final-state particles composed of $\ell^+\ell^-\gamma$ and $\ell^+\ell^-\gamma\gamma$.}
  \label{fit_result} 
\end{figure*}

Figure~\ref{fit_result} shows the $M^{\rm 3C}(\gamma_{2}\ell^{+}\ell^{-})$ distributions for selected candidates of the two channels
of $J/\psi\rightarrow e^{+}e^{-}$ and $J/\psi\rightarrow\mu^{+}\mu^{-}$, where clear signals of $\chi_{c0,1,2}$ can be observed.
No evident $\eta_{c}(2S)$ signature is found. A simultaneous unbinned maximum likelihood fit is performed to obtain the signal yields.
The common parameter for the two $J/\psi$ decay channels is the product branching fraction ($\mathcal{B^{\rm product}}$) of the cascade decays $\psi(3686)\to\gamma\chi_{c0,1,2}(\eta_c(2S))\to\gamma\gamma J/\psi$. The number of signal events for each channel is $N^{\psi(3686)}\times \mathcal{B^{\rm product}}\times\mathcal{B}(J/\psi\rightarrow \ell^{+}\ell^{-})\times\epsilon$.
In the fit, the branching fractions
for $J/\psi\rightarrow e^{+}e^{-}/\mu^{+}\mu^{-}$ and the total number of $\psi(3686)$ events are fixed to the values in Refs.~\cite{PDG} and ~\cite{psip_N}, respectively. The efficiency $\epsilon$ is obtained from the signal MC sample with the higher-order multipole amplitudes considered as listed in Table~\ref{eff}.
The fit contains three $\chi_{c0,1,2}$ components, the $\eta_{c}(2S)$, and the background.
The signal line shapes of the $\chi_{c0,1,2}$ are parametrized as
\begin{equation}\label{chicj_line}
(E_{\gamma 1}^{3}\times E_{\gamma 2}^{3}\times (BW(m)\otimes R\times\epsilon(m)))\otimes G(\mu,\sigma),
\end{equation}
\noindent where $BW(m)$ is the Breit-Wigner function for $\chi_{c0,1,2}$ with the masses and widths fixed at their world average values~\cite{PDG}.
$R$ represents the mass resolution, and $\epsilon(m)$ is the mass-dependent efficiency.
The product [$BW(m)\otimes R\times\epsilon(m)$] can be directly determined from the MC simulation, where the MC events are generated with the simple Breit-Wigner function using the higher-order multipole amplitudes with the angular distributions of the final-state particles. $E_{\gamma 1}$ is the energy of the radiative photon
$\gamma_{1}$ of $\psi(3686)\rightarrow\gamma_{1}\chi_{cJ}$ in the $\psi(3686)$ rest frame, and $E_{\gamma 2}$ is the energy of the $\gamma_{2}$ of
$\chi_{cJ}\rightarrow\gamma_{2}J/\psi$ in the $\chi_{cJ}$ rest frame. The factor $E_{\gamma 1,2}^{3}$ stems from the two-body {\sc PHSP} and the
E1-transition factor, and the Breit-Wigner function modified by the $E_{\gamma 1,2}^{3}$ factor is for the $\chi_{cJ}$ invariant-mass distribution. The line shape is convoluted with a Gaussian function (denoted as $G$) accounting for differences in the invariant mass and mass resolution between the data and the MC simulation.
The mean $\mu$ and standard deviation $\sigma$ of the Gaussian functions are obtained from the fit to the data in a region of [$3.36 < M^{\rm 3C}(\gamma_{2}\ell^{+}\ell^{-}) < 3.61$ GeV/$c^{2}$] by assuming no dependence between the $e^+e^-$ and $\mu^+\mu^-$ decay modes as well as between $\chi_{c0,1,2}$. The results indicate $\mu\leq$ 0.35 MeV/$c^2$ and $\sigma\leq$ 0.73 MeV/$c^2$.
Similarly, the signal line shape of the $\eta_{c}(2S)$ is described by
\begin{equation}\label{etacp_line}
(E_{\gamma 1}^{3}\times E_{\gamma 2}^{7}\times (B(m)\otimes R\times\epsilon(m)))\otimes G(\mu,\sigma),
\end{equation}
\noindent where $E_{\gamma 1}^{3}$ represents the two-body {\sc PHSP} and the M1-transition factor for
$\psi(3686)\rightarrow\gamma_{1}\eta_{c}(2S)$ and $E_{\gamma 2}^{7}$ is the two-body {\sc PHSP} and hindered M1 transition
factor~\cite{Theory_1,Theory_2} for $\eta_{c}(2S)\rightarrow\gamma_{2}J/\psi$. The [$B(m)\otimes R\times\epsilon(m)$] is also determined by MC simulation with the mass and width of $\eta_{c}(2S)$ set to the
world average values~\cite{PDG}. Since the mass of $\eta_{c}(2S)$ is close to those of $\chi_{cJ}$,
the $\mu$ and $\sigma$ of the Gaussian are fixed to the values obtained from a fit to the $\chi_{c0,1,2}$ signals only.

The shapes of backgrounds $\psi(3686)\to\pi^{0}J/\psi,\pi^{0}\pi^{0}J/\psi,\gamma\gamma J/\psi$ and
$e^{+}e^{-}(\to\psi(3686))\rightarrow \ell^{+}\ell^{-}\gamma_{\rm ISR/FSR}(\gamma_{\rm ISR/FSR})$ are taken from MC simulations.
The numbers of $\psi(3686)\rightarrow\pi^{0}J/\psi$ and $\psi(3686)\rightarrow\pi^{0}\pi^{0}J/\psi$ events are fixed to the expectations as given in Table~\ref{eff}.
For the background from $e^{+}e^{-}(\to\psi(3686))\rightarrow \ell^{+}\ell^{-} n\gamma_{\rm ISR/FSR}$, the ratios of
$N_{\ell^{+}\ell^{-}\gamma\gamma}/N_{\ell^{+}\ell^{-}\gamma}$ are fixed to 1.203 for the $e^+e^-$ channel and to 0.689 for the $\mu^+\mu^-$ channel as described above.
In the fit for the final results in the region ($3.36 < M^{\rm 3C}(\gamma_{2}\ell^{+}\ell^{-}) < 3.71$ GeV/$c^{2}$), the parameters of the smearing Gaussians for $\chi_{c0,1,2}$ and $\eta_{c}(2S)$ are fixed, while the numbers of events for $\chi_{c0,1,2}$
and $\eta_{c}(2S)$, $\psi(3686)\rightarrow\gamma\gamma J/\psi$, $e^{+}e^{-}\rightarrow \ell^{+}\ell^{-}\gamma_{\rm ISR/FSR}(\gamma_{\rm ISR/FSR})$
are free parameters. Figure~\ref{fit_result} shows the $M^{\rm 3C}(\gamma_{2}\ell^{+}\ell^{-})$ distributions,
the results of the unbinned maximum likelihood fit, and the relative residuals. The $\chi^{2}/ndf$ of the fit is 1.88 for the $\mu^{+}\mu^{-}$ channel and 1.83
for the $e^{+}e^{-}$ channel.

The product branching fractions from the fit are $(15.8\pm0.3)\times 10^{-4}$, $(351.8\pm1.0)\times 10^{-4}$ and $(199.6\pm0.8)\times 10^{-4}$ for $\chi_{c0,1,2}$ with statistical uncertainty only, respectively.  The branching fraction of $\psi(3686)\to\gamma\gamma J/\psi$ is determined to be $(3.2\pm0.6)\times 10^{-4}$.  All measured branching fractions are consistent with the previous measurement of BESIII~\cite{BESIII_chicj}.
Since no significant $\eta_{c}(2S)$ signal is found, an upper limit at the 90\% C.L. on the product branching fraction is determined by a Bayesian approach using a uniform prior, i.e., finding the values corresponding to 90\% of the probability distribution in the positive domain.

\section{Systematic uncertainties}\label{sys_error_section}

The main sources of systematic uncertainty for the measurements of higher-order multipole amplitudes are the uncertainties in the efficiency,
the kinematic fit procedure, the fit procedure of the combined angular distributions,
statistical fluctuations of the MC sample, and the background contamination.

A simulated sample of events distributed uniformly in {\sc PHSP} is used to normalize the function $W_{\chi_{c1,2}}$. A difference of detection efficiencies between the MC sample and the data will result in a shift in the measurement, which is taken as the systematic uncertainty.
From the studies of the tracking efficiency for electrons and muons with the control samples of
$\psi(3686)\rightarrow\pi^{+}\pi^{-}J/\psi, J/\psi\rightarrow e^{+}e^{-}/\mu^{+}\mu^{-}$ decays, and the photon efficiency with the
control samples from $\psi(3686)\rightarrow 2(\pi^{+}\pi^{-})\pi^{0}$ decays and radiative dimuon events, the difference in the detection
efficiencies between the data and MC is found to be polar angle dependent with the largest value $0.006\pm0.003$, which may change the helicity angular distribution.
The corresponding effect on the higher-order multipole measurement is estimated by varying the efficiency with an asymmetric function of $\cos\theta_{\ell^+}$ and $\cos\theta_{\gamma1}$ as $p(\cos\theta_{\gamma1},\cos\theta_{\ell^{+}})=(1.0+0.003\cos\theta_{\gamma1}-0.006\cos^{2}\theta_{\gamma1})
\times(1.0+0.003\cos\theta_{\ell^{+}}-0.006\cos^{2}\theta_{\ell^{+}})$ [which corresponds to a 0.9\% (0.3\%) difference for $\cos\theta = -1$ (1);
$\theta_{\gamma1}$ is the polar angle for one photon, and $\theta_{\ell^{+}}$ is for one charged track].
Twice the difference with respect to the nominal result is taken as a systematic uncertainty.
For the kinematic fit, the track helix parameters are corrected to reduce the difference in the $\chi_{4\rm C}^{2}$ distribution between the data and
the MC simulation according to the procedure described in Refs.~\cite{GYP,WZHH}. These {\sc PHSP} MC samples without and with the helix correction are used to
normalize $W_{\chi_{c1,2}}$, respectively, and the resultant difference is taken as the systematic uncertainty.

To estimate the uncertainty from the fit procedure, 200 MC samples using the high-order multipole amplitudes are generated, followed by a complete detector simulation. Each sample has 165~thousand (90~thousand) selected events for $\chi_{c1}(\chi_{c2})$, and the same multipole analysis procedure is applied for each sample.
The differences in $a_2^{1}, ~b_2^{1}$ ($a_2^{2},~a_3^{2}, ~b_2^{2},~b_3^{2}$) between the input and fitted values are Gaussian distributed.
The mean values of the Gaussians are $\mu_{a_2^{1}}=(2\pm3)\times 10^{-4}$, $\mu_{b_2^{1}}=(-6\pm3)\times 10^{-4}$ ($\mu_{a_2^{2}}=(17\pm13)\times 10^{-4},~\mu_{a_3^{2}}=(-4\pm8)\times 10^{-4}, ~\mu_{b_2^{2}}=(16\pm6)\times10^{-4},~\mu_{b_3^{2}}=(-32\pm7)\times10^{-4}$) and are taken as the systematic uncertainty.
The statistics of the MC sample for the normalization, about 3.6 (1.8) million events, may affect the fit results. For the normalization function, Eq.(\ref{normal_func}),
the variance for $\overline{a_{n}}(^{n=1,\ldots,9~{\rm for}~\chi_{c1}}_{n=1,\ldots,36~{\rm for}~\chi_{c2}})$ is
\begin{center}\label{square_de_a1}
  $V(\overline{a_{n}}) = \frac{1}{N}\{\frac{\Sigma^{N}_{i=1}a_{n}^{2}(i)}{N}-[\frac{\Sigma^{N}_{i=1}a_{n}(i)}{N}]^{2}\}.$
\end{center}
\noindent The standard deviation for each coefficient is $\sigma(\overline{a_{n}}) = \sqrt{V(\overline{a_{n}})}$. The largest change in parameters
$a_2^{1}$ and $b_2^{1}$ by varying the coefficient by $\pm1\sigma$ for the $\chi_{c1}$ channel ($a_2^{2},~a_3^{2}$, and
$b_2^{2},~b_3^{2}$ for the $\chi_{c2}$ channel) is taken as the systematic uncertainty.

The main backgrounds for the $\chi_{c1}$ channel come from $\psi(3686)\rightarrow\gamma\chi_{c0},\gamma\chi_{c2},\pi^{0}\pi^{0}J/\psi,\gamma\gamma J/\psi$,
which contribute about 0.7\% of the candidates according to a MC study. For the $\chi_{c2}$ channel, the main backgrounds come
from $\psi(3686)\rightarrow\gamma\chi_{c0},\gamma\chi_{c1},\pi^{0}\pi^{0}J/\psi,\gamma\gamma J/\psi$, and
the contribution is about 1\%. In the nominal fit, the contribution of background is estimated by the inclusive MC samples.
To estimate the systematic uncertainty, high-statistics MC samples for backgrounds are generated to redetermine the shape and the contribution according to
previous measurements~\cite{pi0j,PDG,BESIII_chicj}. The difference in the fit results is taken as the systematic uncertainty.
All the systematic uncertainties are summarized in Table~\ref{chicJ_sys}. The total systematic uncertainties are calculated by adding the individual values in
quadrature, thereby assuming that they are independent.

\begin{table*}[!htbp]
\centering
\caption{\label{chicJ_sys}The different sources of systematic uncertainties for the measurement of higher-order multipole amplitudes
for the $\chi_{c1,2}$ channels.}
\begin{tabular}{l|cc|cc|cc}
\hline
    \multirow{2}{*}{Source}    &   \multicolumn{2}{c|}{$\chi_{c1}$} & \multicolumn{4}{c}{$\chi_{c2}$}\\
\cline{2-7}
                               & $a_2^{1}(\times10^{-4})$& $b_2^{1}(\times10^{-4})$& $a_2^{2}(\times10^{-4})$   &     $b_2^{2}(\times10^{-4})$    & $a_3^{2}(\times10^{-4})$  & $b_3^{2}(\times10^{-4})$         \\
\hline
  Efficiency of {\sc PHSP} MC  & 17& 14 &  2  &    4    &  27   &   18         \\
  Kinematic fit                & 8 & 12 &  20 &    9    &  10   &   3          \\
  Fitting procedure            & 2 & 6  &  17 &    16   &  4    &   32         \\
  Statistics of {\sc PHSP} MC  & 2 & 3  &  4  &    2    &  3    &   4          \\
  Background                   & 28& 18 &  23 &    4    &  26   &   4          \\
\hline
  Total                        & 34& 27 &  36 &    20   &   40  &   38         \\
\hline
\end{tabular}
\end{table*}

The systematic uncertainties of the branching fractions measurement include uncertainties from the number of $\psi(3686)$ events (0.9\%)~\cite{psip_N}, the tracking efficiency (0.1\% per lepton)~\cite{sys_tracking}, the
photon detection efficiency (1.0\% per photon)~\cite{sys_photon},
the kinematic fit, the $J/\psi$ mass window, the other selection criteria ($N_{\gamma} \leq 4$, veto $\pi^{0}$ and $\eta$, particle identification, $\cos\theta_{e^+}<0.3 \&\& \cos\theta_{e^-}>-0.3$), the
branching fraction of $J/\psi \rightarrow e^{+}e^{-}/ \mu^{+}\mu^{-}$ (0.6\%)~\cite{PDG}, the interference between $\psi(3686)\to\chi_{c0}\to\gamma\gamma J/\psi$ and nonresonant $\psi(3686)\to\gamma\gamma J/\psi$ processes, and the fitting procedure.

The uncertainty from the kinematic fit is estimated by the same procedure as described in the multipole amplitude measurements.
To estimate the uncertainty caused by the $J/\psi$ mass requirement, a control sample in the
$\chi_{c1,2}$ region $3.49 < M^{4\rm C}(\gamma \ell^+\ell^-) < 3.58$ GeV/$c^{2}$ is used. For data, the only background is from $\psi(3686)\to\pi^0\pi^0J/\psi$, which is determined in fitting with the exclusive MC shape. The efficiency of selection $M^{\rm 4C}(\ell^+\ell^-)\in$(3.08,3.12) GeV is evaluated by comparing the number of signal events before and after the requirement, and the corresponding difference between the data and MC sample is 0.6\% for the $e^{+}e^{-}$ channel and 0.1\% for the $\mu^{+}\mu^{-}$ channel. To be conservative, we take 0.6\% as the systematic uncertainty.
With the same sample, the systematic uncertainties related to the selection
criteria $N_{\gamma} \leq 4$, $\pi^{0}$ veto, $\eta$ veto, and leptons identification are also determined. The overall difference in the efficiency between the
data and MC sample for these criteria is 1.6\% and is taken as a systematic uncertainty. The additional systematic uncertainty due to the polar angle selection for the $e^+e^-$ channel is determined by varying the selection with $\pm0.05$ and fitting simultaneously again. The largest changes on the fit results are taken as the systematic uncertainty.

To estimate the possible uncertainty from the interference between $\psi(3686)\to\gamma \gamma J/\psi$ and $\psi(3686)\to\gamma \chi_{c0}\to\gamma\gamma J/\psi$, we repeat the simultaneous fit, taking the interference into account.  The interference phase is found to be $1.58\pm0.05$. The changes in the signal yields are taken as the systematic uncertainty. Since the signal shapes are determined from MC simulation, the corresponding systematic uncertainty is estimated by an alternative fit with
varying the mass and width of $\chi_{c0,1,2}$ with $\pm 1\sigma$ of the world average values~\cite{PDG} for the signal MC shape.
To estimate the uncertainty due to the background of $\psi(3686)\to\pi^{0}J/\psi, \pi^{0}\pi^{0}J/\psi$ and the ratio
of $N_{\ell^{+}\ell^{-}\gamma\gamma}/N_{\ell^{+}\ell^{-}\gamma}$ for Bhabha and dimuon backgrounds, alternative fits are performed in which the numbers of expected background events (see Table~\ref{eff}) and the ratio of $N_{\gamma\gamma \ell^+\ell^-}/N_{\gamma \ell^+\ell^-}$ are varied by $\pm 1\sigma$. For $\chi_{c0,1,2}$, the largest differences in the signal yields from the nominal values are taken as the systematic uncertainty.
For the $\eta_c(2S)$ case, to be conservative, the one corresponding to the largest upper limit is taken as the final result.
All systematic uncertainties of the different sources are summarized in Table~\ref{sys_all}.
The total systematic uncertainties are obtained by adding the individual ones in quadrature, thereby assuming all these sources are independent.

\begin{table}[!htbp]
\centering
\caption{\label{sys_all}Summary of all systematic uncertainties for the branching fractions measurement.}
\begin{tabular}{lccccc}
\hline
  Source               & $\chi_{c0}$ (\%) & $\chi_{c1}$ (\%) & $\chi_{c2}$ (\%) & $\eta_{c}(2S)$ (\%) \\
\hline
$N_{\psi(3686)}$       &  0.9             &  0.9             &  0.9             &  0.9         \\
Tracking efficiency    &  0.2             &  0.2             &  0.2             &  0.2         \\
Photon detection       &  2.0             &  2.0             &  2.0             &  2.0         \\
Kinematic fit          &  0.6             &  0.5             &  0.5             &  0.4         \\
$J/\psi$ mass window   &  0.6             &  0.6             &  0.6             &  0.6         \\
Other selection        &  2.4             &  2.2             &  2.3             &  2.4         \\
$\mathcal{B}(J/\psi\to e^{+}e^{-}/\mu^{+}\mu^{-})$ &  0.6  &  0.6  &  0.6       &  0.6         \\
Interference           &  0.7             &  -               &  -               &  -           \\
Signal shape           &  0.7             &  0.9             &  1.0             &  -           \\
Background             &  0.1             &  0.1             &  0.1             &  -           \\
\hline
Total                  &  3.6             &  3.4             &  3.5             &  3.4         \\
\hline
\end{tabular}
\end{table}

\section{Result and Summary}

\begin{figure*}[!htpb]
  \centering
    \includegraphics[width=0.8\textwidth]{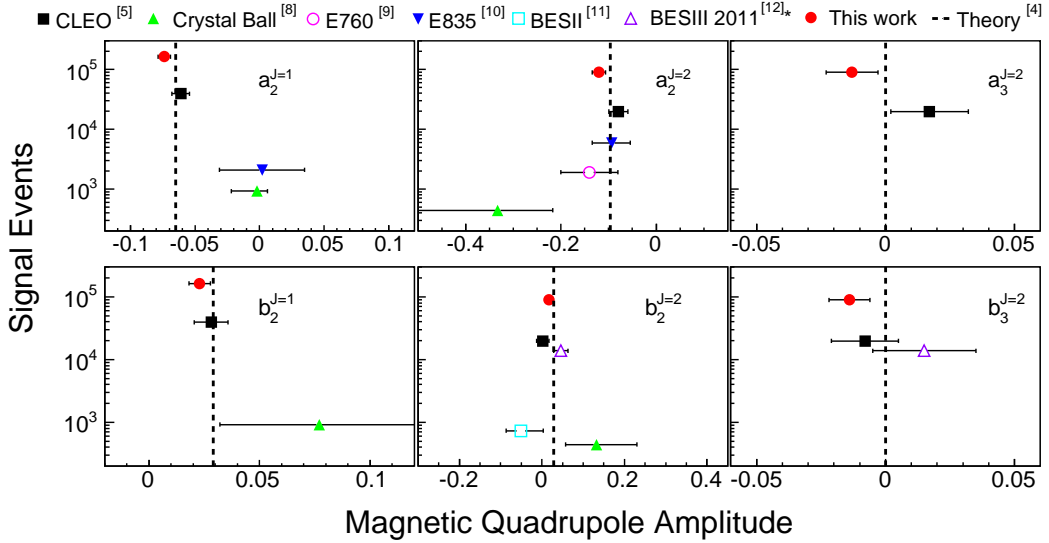}
  \caption{Normalized M2 and E3 amplitudes from this analysis compared with previous experimental results and theoretical predictions~\cite{M2_theory4} with $m_{c} = 1.5$~GeV/$c^2$ and $\kappa = 0$. The $y$ axis shows the number of signal events of each experiment. *Measured by the process of $\psi(3686)\to\gamma\chi_{c2}$ with $\chi_{c2}\to\pi^+\pi^-/K^+K^-$.}
  \label{com_chic12} 
\end{figure*}

Based on 106~million $\psi(3686)$ decays, we measure the higher-order multipole amplitudes for the decays
$\psi(3686)\rightarrow\gamma_{1}\chi_{c1,2}\rightarrow\gamma_1\gamma_{2}J/\psi$ channels. The statistical significance of
nonpure E1 transition is 24.3$\sigma$ and 13.4$\sigma$ for the $\chi_{c1}$ and $\chi_{c2}$ channels, respectively. The normalized M2 contribution for
$\chi_{c1,2}$ and the normalized E3 contributions for $\chi_{c2}$ are listed in Table~\ref{chic1_result_sys}.
Figure~\ref{com_chic12} shows a comparison of our results with previously published measurements and with theoretical predictions with $m_{c} = 1.5$~GeV/$c^2$ and $\kappa = 0$.
The results are consistent with and more precise than
those obtained by CLEO-c~\cite{CLEO-c} and confirm theoretical predictions~\cite{M2_theory1,M2_theory2}. The M2 contributions for
$\psi(3686)\rightarrow\gamma_{1}\chi_{c1}~(b_{2}^{1})$, $\chi_{c1}\rightarrow\gamma_{2}J/\psi~(a_{2}^{1})$, and
$\chi_{c2}\rightarrow\gamma_{2}J/\psi~(a_{2}^{2})$ are found to be significantly nonzero.
The ratios of M2 contributions of $\chi_{c1}$ to $\chi_{c2}$ are
independent of the mass $m_{c}$ and the anomalous magnetic moment
$\kappa$ of the charm quark at leading order in $E_{\gamma}/m_{c}$.  They are determined to be
\begin{equation}\label{ratio_M2}
\begin{split}
&b_{2}^{1}/b_{2}^{2} = 1.35\pm0.72,\\
&a_{2}^{1}/a_{2}^{2} = 0.617\pm0.083.\\
\end{split}
\end{equation}

\noindent The corresponding theory predictions are $(b_{2}^{1}/b_{2}^{2})_{\rm th} = 1.000\pm0.015$ and
$(a_{2}^{1}/a_{2}^{2})_{\rm th} = 0.676\pm0.071$~\cite{CLEO-c}.
By using the most precise measurement of the M2 amplitudes $a_{2}^{1}$ and by taking $m_{c}=1.5\pm0.3$~GeV/$c^2$, the anomalous magnetic moment
$\kappa$ can be obtained
from Eq.~(\ref{theory_predict}),
\begin{equation}\label{kappa}
\begin{split}
1+\kappa = &-\frac{4m_{c}}{E_{\gamma_{2}}[\chi_{c1}\rightarrow\gamma_{2}J/\psi]}a_{2}^{1}\\
 =& 1.140\pm0.051\pm0.053\pm0.229,\\
\end{split}
\end{equation}
\noindent where the first uncertainty is statistical, the second uncertainty is systematic, and the third uncertainty is from $m_{c}=1.5\pm0.3$~GeV$/c^2$.

\begin{figure*}[!htbp]
  \centering
  \includegraphics[width=0.6\textwidth]{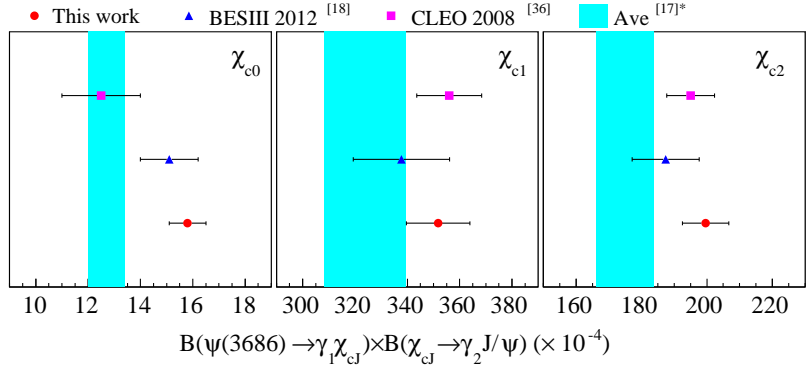}\\
  \caption{Comparison of the product branching fraction $\mathcal{B}(\psi(3686)\rightarrow\gamma_1 \chi_{cJ})\times\mathcal{B}(\chi_{cJ}\rightarrow\gamma_2 J/\psi)$ with previously published measurements. *The average "Ave" is the product between the individual world average of $\mathcal{B}(\psi(3686)\rightarrow\gamma_1 \chi_{cJ})$~\cite{PDG} and  $\mathcal{B}(\chi_{cJ}\rightarrow\gamma_2 J/\psi)$~\cite{PDG}.}
  \label{final_result_compare}
\end{figure*}

Based on the multipole analysis, we measure the product branching fractions for $\psi(3686)\rightarrow\gamma\chi_{c0,1,2}\to\gamma\gamma J/\psi$ to be $(15.8\pm0.3\pm0.6)\times 10^{-4}$, $(351.8\pm1.0\pm12.0)\times 10^{-4}$, and $(199.6\pm0.8\pm7.0)\times 10^{-4}$, respectively, where the first uncertainty is statistical and the second is systematic. In Fig.~\ref{final_result_compare}, the product branching fractions are compared to previous results from BESIII~\cite{BESIII_chicj}, CLEO~\cite{CLEO_chicj}, and the world average~\cite{PDG}. The world average refers to the product of the average branching fraction of $\psi(3686)\rightarrow\gamma_{1}\chi_{cJ}$ and the average branching fraction of $\chi_{cJ}\rightarrow\gamma_{2} J/\psi$, where the results of BESIII and CLEO are not included in the world average values. For all $\chi_{cJ}$, our results exceed the precision of the previous measurements. Compared to the previous BESIII result, the results are consistent within 1$\sigma$, but we have considered the higher-order multipole amplitudes and improved the systematic uncertainty due to a more precise measurement of the total number of produced $\psi(3686)$~\cite{psip_N}. In addition, our measurement for the $\chi_{c0}$ channel is 3$\sigma$ larger than the result from CLEO and 3$\sigma$ larger than the world average value, while for the $\chi_{c1,2}$, our results are consistent with previous measurements.
There are theoretical predictions for the branching fraction $\psi(3686)\to \gamma\chi_{c0,1,2}$ by several different models~\cite{QM_1,QM_chi1,Theory_1} without consideration of higher-order multipole amplitudes, which agree with each other poorly.
The results in this measurement will provide a guidance for the theoretical calculations.

We also search for the decay $\eta_{c}(2S)\rightarrow\gamma J/\psi$ through $\psi(3686)\rightarrow\gamma\eta_{c}(2S)$.
No statistically significant signal is observed. Considering the systematic uncertainty,
an upper limit on the product branching fraction is determined to be $\mathcal{B}(\psi(3686)\rightarrow\gamma\eta_{c}(2S))\times\mathcal{B}(\eta_{c}(2S)\rightarrow\gamma J/\psi) < 9.7\times 10^{-6}$ at the 90\% C.L., where the systematic uncertainty is incorporated by a factor $1/(1-\sigma_{\rm syst.})$ for conservative.
Combining the result of $B(\psi(3686)\to\gamma\eta_c(2S))$ obtained by BESIII~\cite{wangll}, the upper limit of the branching fraction for $\eta_c(2S)\to\gamma J/\psi$ is
$\mathcal{B}(\eta_{c}(2S)\rightarrow\gamma J/\psi) < 0.044$ at the 90\% C.L. Using the width of $\eta_c(2S)$ of $11.3^{+3.2}_{-2.9}$~MeV/$c^2$~\cite{PDG}, our upper limit
implies a partial width of $\Gamma(\eta_{c}(2S)\rightarrow\gamma J/\psi) <$ 0.50~MeV/$c^2$. Although this result agrees with the prediction of LQCD (0.0013 MeV/$c^2$)~\cite{LQCDpredict2},
it clearly has a very limited sensitivity to rigorously test the theory.

\acknowledgements

The BESIII Collaboration thanks the staff of BEPCII and the Institute of High Energy Physics computing center for their strong support. This work is supported in part by National Key Basic Research Program of China under Contract No. 2015CB856700; National Natural Science Foundation of China (NSFC) under Contracts No. 11125525, No. 11235011, No. 11322544, No. 11335008, No. 11425524, No. 11475187, No. 11521505, and No. 11575198; the Chinese Academy of Sciences (CAS) Large-Scale Scientific Facility Program; the CAS Center for Excellence in Particle Physics; the Collaborative Innovation Center for Particles and Interactions; Joint Large-Scale Scientific Facility Funds of the NSFC and CAS under Contracts No. 11179007, No. U1232201, and No. U1332201; CAS under Contracts No. KJCX2-YW-N29 and No. KJCX2-YW-N45; 100 Talents Program of CAS; National 1000 Talents Program of China; Institute of Nuclear and Particle Physics and Shanghai Key Laboratory for Particle Physics and Cosmology; German Research Foundation under Collaborative Research Center Contract No. CRC-1044; Istituto Nazionale di Fisica Nucleare, Italy; Koninklijke Nederlandse Akademie van Wetenschappen under Contract No. 530-4CDP03; Ministry of Development of Turkey under Contract No. DPT2006K-120470; NSFC under Contracts No. 11405046 and No. U1332103; Russian Foundation for Basic Research under Contract No. 14-07-91152; The Swedish Research Council; U. S. Department of Energy under Contracts No. DE-FG02-04ER41291, No. DE-FG02-05ER41374, No. DE-SC0012069, No. DESC0010118; U.S. National Science Foundation; University of Groningen and the Helmholtzzentrum fuer Schwerionenforschung GmbH, Darmstadt; and World-Class University Program of National Research Foundation of Korea under Contract No. R32-2008-000-10155-0.

\end{document}

%% file: author_20151124_new.tex
\author{
\begin{small}
\begin{center}
M.~Ablikim$^{1}$, M.~N.~Achasov$^{9,e}$, X.~C.~Ai$^{1}$, O.~Albayrak$^{5}$, M.~Albrecht$^{4}$, D.~J.~Ambrose$^{44}$, A.~Amoroso$^{49A,49C}$, F.~F.~An$^{1}$, Q.~An$^{46,a}$, J.~Z.~Bai$^{1}$
% ...
\\
\vspace{0.2cm}
(BESIII Collaboration)\\
\vspace{0.2cm} {\it
$^{1}$ Institute of High Energy Physics, Beijing 100049, People's Republic of China\\
% ...
$^{4}$ Bochum Ruhr-University, D-44780 Bochum, Germany\\
$^{5}$ Carnegie Mellon University, Pittsburgh, Pennsylvania 15213, USA\\
% ...
$^{9}$ G.I. Budker Institute of Nuclear Physics SB RAS (BINP), Novosibirsk 630090, Russia\\
% ...
$^{44}$ University of Rochester, Rochester, New York 14627, USA\\
% ...
$^{46}$ University of Science and Technology of China, Hefei 230026, People's Republic of China\\
% ...
$^{49}$ (A)University of Turin, I-10125, Turin, Italy; (B)University of Eastern Piedmont, I-15121, Alessandria, Italy; (C)INFN, I-10125, Turin, Italy\\
% ...
$^{a}$ Also at State Key Laboratory of Particle Detection and Electronics, Beijing 100049, Hefei 230026, People's Republic of China\\
% ...
$^{e}$ Also at the Novosibirsk State University, Novosibirsk, 630090, Russia\\
% ...
}\end{center}
\vspace{0.4cm}
\end{small}
}

%% file: etacp_gammajpsi.bbl
\begin{thebibliography}{99}
\bibitem{M2_theory1}J. L. Rosner, Phys. Rev. D \textbf{78}, 114011 (2008).
\bibitem{M2_theory2}K. J. Sebastian, H. Grotch and F. L. Ridener, Phys. Rev. D \textbf{45}, 3163 (1992).
\bibitem{M2_theory3}P. Moxhay and J. L. Rosner, Phys. Rev. D \textbf{28}, 1132 (1983).
\bibitem{M2_theory4}G. Karl {\it et al.}, Phys. Rev. Lett. \textbf{45}, 215 (1980).
\bibitem{CLEO-c}M. Artuso {\it et al.} (CLEO Collaboration), Phys. Rev. D \textbf{80}, 112003 (2009).
\bibitem{angular1}P. K. Kabir and A. J. G. Hey, Phy. Rev. D \textbf{13}, 3161 (1976).
\bibitem{angular2}C. Edwards {\it et al.}, Phys. Rev. D \textbf{25}, 3065 (1982).
\bibitem{crystal_ball}M. Oreglia {\it et al.} (Crystal Ball Collaboration), Phys. Rev. D \textbf{25}, 2259 (1982).
\bibitem{E-760}T. Armstrong {\it et al.} (E760 Collaboration), Phys. Rev. D \textbf{48}, 3037 (1993).
\bibitem{E-835}M. Ambrogiani {\it et al.} (E835 Collaboration), Phys. Rev. D \textbf{65}, 052002 (2002).
\bibitem{BESII}M. Ablikim {\it et al.} (BES Collaboration), Phys. Rev. D \textbf{70}, 092004 (2004).
\bibitem{BESIII}M. Ablikim {\it et al.} (BESIII Collaboration), Phys. Rev. D \textbf{84}, 092006 (2011).
\bibitem{QM}S. Godfrey and N. Isgur, Phys. Rev. D \textbf{32}, 189 (1985).
\bibitem{QM_1}E. Eichten {\it et al.} Phys. Rev. D \textbf{21}, 203 (1980).
\bibitem{QM_chi1}N. Brambilla {\it et al.} (QWG Collaboration), arXiv:hep-ph/0412158.
\bibitem{Theory_1}T. Barnes, S. Godfrey and E. S. Swanson, Phys. Rev. D \textbf{72}, 054026 (2005).
\bibitem{PDG}K. A. Olive {\it et al.} (Particle Data Group), Chin. Phys. C \textbf{38}, 090001 (2014).
\bibitem{BESIII_chicj}M. Ablikim {\it et al.} (BESIII Collaboration), Phys. Rev. Lett. \textbf{109}, 172002 (2012).
\bibitem{psip_N}M. Ablikim {\it et al.} (BESIII Collaboration), Chin. Phys. C \textbf{37}, 063001 (2013).
\bibitem{3773_N}M. Ablikim {\it et al.} (BESIII Collaboration), Chin. Phys. C \textbf{37}, 123001 (2013).
\bibitem{BESIII_detector}M. Ablikim {\it et al.} (BESIII Collaboration), Nucl. Instrum. Methods Phys. Res, Sect. A \textbf{614}, 345 (2010).
\bibitem{photon_resolution}M. Ablikim {\it et al.} (BESIII Collaboration), Phys. Rev. D \textbf{85}, 112008 (2012).
\bibitem{GEANT4}S. Agostinelli {\it et al.} (GEANT4 Collaboration), Nucl. Instrum. Methods Phys. Res, Sect. A \textbf{506}, 250 (2003).
\bibitem{KKMC}S. Jadach, B. F. L. Ward, and Z. Was, Comput. Phys. Commun. \textbf{130}, 260 (2000); Phys. Rev. D \textbf{63}, 113009 (2001).
\bibitem{EvtGen}D. J. Lange, Nucl. Instrum. Meth. A \textbf{462} 152 (2001); R. G. Ping, Chin. Phys. C \textbf{32}, 599 (2008).
\bibitem{LundCharm}J. C. Chen, G. S. Huang, X. R. Qi, D. H. Zhang, and Y. S. Zhu, Phys. Rev. D \textbf{62}, 243 (2000).
\bibitem{babayaga}G.~Balossini, C.~M.~Carloni Calame, G.~Montagna, O.~Nicrosini, and F.~Piccinini, Nucl.\ Phys. \textbf{B758}, 227 (2006).
\bibitem{amp_012}G. Karl {\it et al.} Phys. Rev. D \textbf{13}, 1203 (1976).
\bibitem{Pearson_test}R. L. Plackett, K. Pearson and the Chi-Squared Test, Int. Stat. Rev. \textbf{51}, 59 (1983).
\bibitem{Theory_2}J. J. Dudek, R. Edwards and C. E. Thomas, Phys. Rev. D \textbf{79}, 094504 (2009).
\bibitem{GYP}M. Ablikim {\it et al.} (BESIII Collaboration), Phys. Rev. D \textbf{87}, 012002 (2013).
\bibitem{WZHH}M. Ablikim {\it et al.} (BESIII Collaboration), Phys. Rev. D \textbf{91}, 112005 (2015).
\bibitem{pi0j}M. Ablikim {\it et al.} (BESIII Collaboration), Phys. Rev. D \textbf{86}, 092008 (2012).
\bibitem{sys_tracking}M. Ablikim {\it et al.} (BESIII Collaboration), Phys. Rev. D \textbf{88}, 032007 (2013).
\bibitem{sys_photon}M. Ablikim {\it et al.} (BESIII Collaboration), Phys. Rev. D \textbf{88}, 112001 (2013).
\bibitem{CLEO_chicj}H. Mendez {\it et al.} (CLEO Collaboration), Phys. Rev. D \textbf{78}, 011102 (2008).
\bibitem{wangll}M. Ablikim {\it et al.} (BESIII Collaboration), Phys. Rev. Lett. \textbf{109}, 042003 (2012).
\bibitem{LQCDpredict2}D. Becirevic and F. Sanfilippo,  J. High Energy Phys. 01 (2013) 028.
\end{thebibliography}
